\documentstyle[11pt,newpasp,twoside]{article}
\def\edcomment#1{\iffalse\marginpar{\raggedright\sl#1\/}\else\relax\fi}
\reversemarginpar

\newcommand{\beq}{\begin{equation}}
\newcommand{\eeq}{\end{equation}}
\def\beqa{\begin{eqnarray}}
\def\eeqa{\end{eqnarray}}

\def\lb{\langle}
\def\rb{\rangle}

\begin{document}
\title{The Galaxy and Mass $N$-Point Correlation Functions: a
Blast from the Past} 
\author{P. J. E. Peebles}
\affil{Joseph Henry Laboratories, Princeton University,
Princeton, NJ 08544}

\begin{abstract}
Correlation functions and related statistics have been
favorite measures of the distributions of extragalactic objects
ever since people started analyzing the clustering of the
galaxies in  the 1930s. I review the evolving reasons for this
choice, and comment on some of the present issues in the
application and interpretation of these statistics, with
particular attention to the question of how closely galaxies
trace mass.  
\end{abstract}

\section{Introduction}
The choice of statistical measures of the 
distributions of galaxies and other extragalactic objects depends
on the questions one wants to address, and it is influenced by 
tradition and personal taste. Both have favored correlation
functions and related measures, such as the moments of counts in
cells, for reasons that have evolved with the subject. This
review of the role of $N$-point correlation functions in analyses
of the distributions of galaxies and mass is a folk history, that
reflects my experiences. I offer it as raw material for a real
history of the development of modern cosmology, and, I hope, as a
useful perspective on present directions of research.

With the exception of some basic definitions and relations,
presented in \S 2.1, I refer to references for all technical
details. Almost all I know is in Peebles (1973, hereafter SAC)
and Peebles (1980, hereafter LSS). It used to be good form to
refer to a selection of recent papers through which the reader
could trace earlier contributions. But now even the
more computer-challenged, such as me, can as easily trace forward
in time through citation links, and the weight of selection of
references may be shifted to earlier papers. This informs my 
choice of references to examples of current research.

\section{Two-Point Correlation Functions and Power Spectra}

\subsection{Definitions}

Modern cosmology starts with Einstein's (1917) assumption that
the observable universe is homogeneous and isotropic in the
large-scale average. I believe Jerzy Neyman and Elizabeth Scott
(commencing with Neyman \&\ Scott 1952) were the first
to translate this to the approximation of the galaxy distribution
as a realization of a stationary random point process. At the
time the evidence for homogeneity was sparse; it now seems
compelling (Peebles 1993, \S 3, \S 7; Davis 1997). 

The number density, $n$, in a stationary point process in three
dimensions may be defined by the probability a point appears in
the randomly placed volume element $dV$,   
\beq
dP = n\, dV, 
\eeq
and the two-point correlation function by the probability points
appear in each of the volume elements $dV_1$ and $dV_2$ at
separation $r_{12}$,  
\beq
dP = n^2dV_1dV_2 [1 + \xi (r_{12})].
\label{eq:xi}
\eeq
The analog for a continuous function, $f(\vec r)$, is the
autocorrelation function, 
\beq
\xi _c(r) = \lb f(\vec x + \vec r)f(\vec x)\rb /\lb f\rb ^2 - 1.
\eeq
If $f$ is constant, $\xi _c$ vanishes. In a homogenous Poisson
point process, the probability a point appears in a given volume
element is statistically independent of what happens everywhere
else, so the reduced function, $\xi$, vanishes.

The Fourier amplitude of the distribution of points at positions
$\vec r_i$, in flat space periodic in a box of volume $V_u$, is 
\beq
\delta _{\vec k} = \sum e^{i\vec k\cdot\vec r_i}.
\eeq
The expectation value of the square of the amplitude defines the
power spectrum, 
\beq
P(k) =\int d^3r\xi (r) e^{i\vec k\cdot\vec r} =
{\lb |\delta _{\vec k}|^2\rb\over n^2V_u} - {1\over n}.
\eeq
The $\delta _{\vec k}$ are uncorrelated, 
\beq
\lb\delta _{\vec k}\delta _{\vec k'}\rb = 0\quad\hbox{if}\quad
	\vec k\not=\vec k',
\eeq
though not in general statistically independent. In a fair
sample --- size large compared to the scale over which the
process is statistically related --- the probability
distribution of the power spectrum for each $\vec k$ is
exponential, a consequence of the central limit theorem.

The variance --- the second central moment --- of the count of
points in a randomly placed volume $V$ is
\beq
\sigma ^2 = \lb (N -\lb N\rb )^2\rb = 
nV + n^2\int _{V} dV_1dV_2 \xi (r_{12}), 
\label{eq:variance}
\eeq
in terms of the correlation function, and 
\beq
\sigma ^2 = nV + n^2\int d^3k\, P(k) |W_{\vec k}|^2,\qquad W_{\vec k} 
= \int _{V}d^3r e^{i\vec k\cdot\vec r}/(2\pi )^{3/2},
\label{eq:kvariance}
\eeq
in terms of the power spectrum. 

\subsection{Power Spectra and Correlation Functions}

The Princeton program of
$N$-point correlation function analyses of the distributions of 
extragalactic objects grew out of a question Sydney van den
Bergh, then at the David Dunlap Observatory, the University of
Toronto, put to me in 1966. Does the distribution of Abell's
(1958) rich clusters of galaxies exhibit convergence to the  
large-scale homogeneity assumed in most cosmologies? At the time 
the best evidence for large-scale homogeneity came from Hubble's
deep galaxy counts and the near isotropy of the radio, microwave,
and X-ray radiation backgrounds (Peebles 1971, chapter II). 
Abell's catalog was the deepest large sample of objects
with good distance estimates. There are two parts to
van den Bergh's question. First, might the large fluctuations in
the cluster distribution mean Abell's catalog samples only part
of a clustering hierarchy that extends to larger scales (Kiang
1967; de Vaucouleurs 1970)? Second, might the fluctuations be
only apparent, a result of patchy obscuration 
(Neyman \&\ Scott 1952; Limber 1953)? As will be discussed, both
are tested by the scaling of statistical measures of the
clustering with depth.

Our first choice of statistic was driven by Blackman \&\ Tukey's
(1959) eloquent demonstrations of the advantages of power spectra
over correlation 
functions in detecting a weak signal. For a distribution on the 
sphere one replaces the Fourier transform with a spherical
harmonic transform, $a_l^m$, with power spectrum $|a_l^m|^2$ 
(Yu \&\ Peebles 1969). In a fair sample, where the coherence
length is small compared to the catalog depth, the distribution of
the $|a_l^m|^2$ is close to exponential, as noted in \S 2.1, and
the slope of the exponential distribution is a useful measure of
the number of   
objects per independent clump (Yu \&\ Peebles 1969, eq. [15]).
The coherence length in the Abell sample is not very much smaller 
than the sample depth, but the distributions of the $|a_l^m|^2$
are strikingly close to exponential (Hauser \&\ Peebles 1973,
Fig.~4). The key result, that answered van den Bergh's question, 
is that the power spectrum of the angular distribution of the
clusters scales with depth in the way expected if 
Abell's catalog is a fair sample of a stationary process, as
opposed to a sample from a clustering hierarchy that extends to
larger scales (Hauser \&\ Peebles 1973, Fig. 9). This analysis is
continued in Bahcall \&\ Soneira (1983).     

The advantages of power spectra over correlation functions for
the measurement of small fluctuations make the spectrum the
statistic of choice for analyses of the anisotropy of
the 3K thermal cosmic radiation (the CBR; Hu et al. 2000), and
measurements of large-scale fluctuations in the galaxy
distribution (Peacock \&\ Dodds 1994). These analyses 
usually remove the part of sky at low galactic latitude, where
interference by the Milky Way is most serious. This makes the 
measured spherical harmonic amplitudes convolutions of the
true $a_l^m$ (SAC eq.~[42]). Examples of the resulting
correlation in the power spectrum are in Hauser \&\ Peebles
(1973). The convolution is not a
problem for analyses of catalogs, where the true $|a_l^m|^2$
varies slowly with $l$. But the relatively large dipole term in
the CBR anisotropy, from our peculiar velocity, interferes with
estimates of higher order spherical harmonic amplitudes. 
Wright's (1993) 
remedy is to define new expansion functions that are orthogonal
to the dipole terms in the observed part of the sky. G\'{o}rski
(1994) discusses expansion functions that are orthogonal in the
observed sky, so all the expansion coefficients are uncorrelated
(though not statistically independent). 

The statistics of choice for the measurement of the small-scale
strongly nonlinear fluctuations in the galaxy distribution are
correlation functions, for they are easier to measure and
interpret than their Fourier or spherical harmonic transforms.
This follows the first studies of fluctuations in the galaxy
distribution, that used the variance of galaxy counts in cells
(Bok 1934; Zwicky 1953; Rubin 1954; Neyman, Scott \&\ Shane 1954;
Kiang 1967) The natural step is to the mean lagged product, or
two-point correlation function 
(Limber 1953; Rubin 1954; Neyman, Scott \&\ Shane 1954; Irvine
1961; Layzer 1963; Kiang 1967; Kiang \&\ Saslaw 1969; Totsuji \&\
Kihara 1969). Further comments on the use of galaxy correlation
functions and power spectra are in LSS \S 29.  

\subsection{Scaling} 

In the first galaxy catalogs large enough for statistical
analyses of clustering, the only distance measure was the 
apparent magnitude or magnitude limit, and the luminosity function 
was not well known. But the following considerations allowed
useful measures of clustering and tests for reliability of the
results. 

The (ensemble average) reduced angular correlation function in a
catalog of galaxies 
selected by apparent magnitude is a linear integral over the
reduced spatial function. This very useful relation was derived
by Limber (1953). The integral relation depends
on a selection function --- the fraction of galaxies selected for
the catalog as a function of distance --- and the selection
function depends on the galaxy luminosity function.  

If the catalog is close to a fair sample the reduced angular
correlation functions are substantially different from zero only
at angles $\theta\ll 1$. Then the scaling of the angular
statistic with apparent magnitude is independent of the
selection function (provided the selection function scales with
apparent magnitude according to the inverse square law; SAC
eqs.~[67] and [69]). If the estimates agree with this scaling it
argues the catalog is close to a fair sample, not seriously
affected by variable obscuration.  

Scaling in Abell's (1958) catalog of clusters of
galaxies is demonstrated in Hauser \&\ Peebles (1973), as
noted above. Scaling for galaxies is demonstrated in Peebles
\&\ Hauser (1974) and Groth \&\ Peebles (1977) for the 
Zwicky et al. (1961-68) 
catalog, the deeper Lick Catalog (Shane \&\ Wirtanen 1967), and
the still deeper Jagellonian Field (Rudnicki et al. 1973). 
Maddox et al. (1990) extend the demonstration to the
APM sample, at about the same depth as the Jagellonian sample,
but in a much larger field. The scaling is persuasive evidence
that we have a fair measure of the galaxy distribution. 

The present issue of interest is the departure from scaling, at
depths large enough to reveal the evolution of the clustering of
the galaxies (Hogg, Cohen \&\ Blandford 2000, Fig. 7).

\subsection{The Galaxy Correlation Function}

In the scaling limit a power law spatial function,
$\xi (r)\propto r^{-\gamma}$, with $\gamma >1$, produces a power
law angular function, $w(\theta )\propto r^{1-\gamma}$,
independent of the selection function. Totsuji
\&\ Kihara (1969) found the first evidence that the galaxy
function is close to a power law. The evidence is extended, and 
scaling demonstrated, in  Groth \&\ Peebles (1977). The 
power law behavior is extended to still smaller scales in Gott
\&\ Turner (1979), and to larger scales in Maddox et al. (1990), 
showing the spatial function is well approximated
as\footnote{Hubble's constant is 
$H_o=100h$ km s$^{-1}$ Mpc$^{-1}$.}
\beq
\xi (r) = (r_o/r)^\gamma, \qquad 
\gamma = 1.77 \pm 0.04,\qquad hr_o = 5\pm 0.5\hbox{ Mpc},
\label{eq:power_law}
\eeq
at the range of separations 
\beq
\qquad 10\hbox{ kpc}\la hr\la 10\hbox{ Mpc}.
\label{eq:range}
\eeq
At the lower bound the luminous parts of the galaxies
nearly overlap. An extension to still smaller scales, in the
galaxy-mass cross correlation function, is discussed in \S 4.1.  
At $hr\sim 20$~Mpc the galaxy correlation function breaks below
the power law. This appears in the Lick and Jagellonian samples
(Groth \&\ Peebles 1977; Fry \&\ Seldner 1982), but perhaps too
close to the systematic errors to be convincing. The break is
well established in the APM sample. It is thought that at larger
separations the galaxy distribution is anticorrelated, $\xi <0$. 
This would mean that at small wavenumber, $k$, the power 
spectrum, $P(k)$, increases with increasing $k$. Detecting this 
requires a relatively deep sample and good control of the 
selection function as a function of position in the sky. The 
effect likely is seen (Sutherland et al. 1999, Fig. 9), in
measurements of $P(k)$ that extend to $hr\sim 100$~Mpc. 

Catalogs in progress will tighten bounds on departures from a 
small-scale power law, including the evidence that
$\xi (r)$ rises above the power law at $r\sim r_o$ before the
break down at $r\sim 3r_o$ (Soneira \&\ Peebles 1978, Fig. 6).
Modern catalogs have distance measures from redshifts and
predictors of luminosities, but neither is accurate enough for a
direct reconstruction of the small-scale spatial distribution.
Analyses will follow Limber (1953) and Rubin (1954) in deriving
spatial functions from projected functions (as in Davis
\&\ Peebles 1983).

\subsection{Peculiar Velocities}

One does need redshifts to measure galaxy peculiar velocities. In 
the space of galaxy redshift and angular
position the two-point correlation function is a function of two 
variables, the transverse and radial separation, the former 
probing spatial clustering and the latter relative peculiar
velocities (Davis \&\ Peebles 1983).  
Large-scale flows are usefully measured by the power
spectrum in redshift space (Kaiser 1987), and by the peculiar
velocity autocorrelation function derived from distance
predictors (Groth, Juszkiewicz \&\ Ostriker 1989). On relatively
small scales the rms relative peculiar velocity is dominated by
the rich clusters (Marzke et al. 1995); here
moments are not the best measure of the distribution of
relative velocities (Davis, Miller \&\ White
1997). 

In \S 4.2 I comment on early advances in the measurements of
galaxy peculiar velocities, in connection with the biased galaxy
formation picture. For the state of the art, see Courteau, Strauss
\&\ Willick (2000).   

\section{Higher Moments and Clustering Models}

One uses statistics to reduce a lot of information to a more 
interesting and understandable quantity. It can be useful to have
a sequence of statistics, that allows recovery of progressively
more detail. Thus one may characterize a single
random number by its mean and second central moment, then the
third moment, and on to higher moments, and one may 
characterize the galaxy distribution by the mean number density,
then the two-point correlation function, then the three-point
function, and so on. 

\subsection{Higher Order Correlation Functions}

Following the definition of the two-point correlation function
for a stationary point process in three dimensions
(eq.~[\ref{eq:xi}]), one writes the probability of finding
points in the three volume elements $dV_1$, $dV_2$ and $dV_3$
that define a triangle with sides $r_{12}$, $r_{23}$ and $r_{31}$
as  
\beq
dP = n^3dV_1dV_2dV_3 [1 + \xi (r_{12}) + \xi (r_{23})
 + \xi (r_{31}) + \zeta (r_{12},r_{23},r_{31})].
\label{eq:zeta}
\eeq
In an idealized ensemble of catalogs of angular positions the
probability of finding points in the elements of solid angle
$d\Omega _1$, $d\Omega _2$ and $d\Omega _3$ is similarly
written 
\beq
dP = {\cal N}^3d\Omega _1d\Omega _2d\Omega _3 
[1 + w(\theta _{12}) + w(\theta _{23})
 + w(\theta _{31}) + z (\theta _{12},\theta _{23},\theta _{31})],
\label{eq:z}
\eeq
where ${\cal N}$ is the mean surface number density. 

The linear combination in brackets in equation~(\ref{eq:z}) has a
simple interpretation: the two-point angular 
function $w(\theta _{12})$ represents the probability points~1
and~2 are correlated in space, with point 3 accidentally close in 
projection, and so on, so that the reduced angular function,  
$z (\theta _{12},\theta _{23},\theta _{31})$, is an integral over
the reduced spatial function, $\zeta (r_{12},r_{23},r_{31})$. 
Since galaxy distances never will be known well enough to
untangle galaxies seen close in projection, it remains very
useful to have this simple way to take account of projections.

At least five considerations motivated the introduction of
the three-point correlation function to the analysis of the
galaxy distribution (Peebles \&\ Groth 1975). First, Saslaw's
(1972) discussion of the theory of gravitational dynamics
demonstrated to us the usefulness of methods from the treatment
of a non-ideal gas. This includes the hierarchy of $N$-point
correlation functions in position and momentum. Second, Munk
(1966) showed a useful application to data, in the behavior of
ocean waves. Munk discussed the third moment of 
the time series of wave height at a fixed position, and the
bispectrum --- the Fourier transform of the three-point
correlation function --- of ocean wave trains. Fry \&\ Seldner
(1982) gave the first estimates of the bispectrum of the galaxy
distribution. The remarkable advances in the analysis of the
bispectrum on the scale of weakly nonlinear fluctuations 
are noted in \S 4.3. Third, the quite stable estimates of the
galaxy two-point function, evidenced by the success of the
scaling test, showed us the catalogs contain other useful
information, that could help distinguish different clustering
models with  the same two-point function. Fourth, I suspected the 
gravitational growth of galaxy clustering would make the 
reduced three-point function vary with the size of the triangle
defined by the three points as the square of the two-point
function (Peebles 1974a). Measurements of the galaxy three-point
function thus might test the gravitational instability picture.
The status of this idea is 
discussed in \S 4. And finally, the first estimates of the
angular three-point function, in Zwicky's (1961-68) catalog, were
seen to be consistent with this conjecture, in the very
convenient form  
\beq
z = q[w(\theta _{12})w(\theta _{23}) + 
w(\theta _{23})w(\theta _{31}) + w(\theta _{31})w(\theta _{12})],
\label{eq:zmodel}
\eeq
where $q$ is a constant. This expression approaches zero when two
points are close and the other far way. The other simple form
constructed from $w(\theta )$ with this wanted property, 
$z\propto w(\theta _{12})w(\theta _{23})w(\theta _{31})$,
is quite inconsistent with the measurements.
The great convenience for the early study of galaxy clustering
is that equation~(\ref{eq:zmodel}) follows from the spatial
function   
\beq
\zeta = Q[\xi (r_{12})\xi (r _{23}) + 
\xi (r _{23})\xi (r _{31}) + \xi (r _{31})\xi (r _{12})].
\label{eq:zetamodel}
\eeq
One only needs the luminosity function or selection
function to translate $q$ to $Q$. 

We introduced equation~(\ref{eq:zetamodel}) to model
the gravitational growth of clustering, and, at least
as important, because it offers a wonderfully easy way to 
translate from the angular to the spatial function. The success
of the scaling test applied to the three-point angular functions
from the Zwicky, Lick, and Jagellonian samples convinced us 
the spatial function is reasonably well measured, and
equation~(\ref{eq:zetamodel}) is a remarkably good approximation. 
With the extension to larger scales, from moments of counts of APM 
galaxies in cells (Gazta\~naga 1994; Szapudi et al. 1995), the
galaxy three-point function is measured to better than a factor
of two at separations in the range
\beq 
0.1\la hr\la 10\hbox{ Mpc}.
\label{eq:zrange}
\eeq
Equation~(\ref{eq:zetamodel}) fits the measured variation of
the three-point correlation function with triangle size, and the
measured variation with triangle shape at $r\la r_o$. In a
remarkable advance, Szapudi et al. (2000b) find that
equation~(\ref{eq:zetamodel}) fits the measured three-point
function back to  redshift $z\sim 1$. 

The four-point function also conveniently approximates products 
of two-point functions (Fry \&\ Peebles 1978). The extension to
the five-point function is an even greater effort (Sharp,
Bonometto \&\ Lucchin 1984). The sensible consensus is
that these higher order correlation functions are best probed by
the moments of counts in cells. But the full three-point
function has proved to be useful, and I would not be surprised to
see a return of interest in the full $N$-point functions at $N=4$
and~5.  

\subsection{The Galaxy Clustering Hierarchy Model}

Neyman and Scott (1952; Neyman 1962) pioneered the
development of analytically prescribed models of the galaxy
distribution. The Neyman-Scott model prescription places galaxies 
in clusters. This model, and a close relative, a halo model that
usefully represents predictions of the adiabatic cold dark matter
(CDM) model for the mass distribution, are discussed in  
\S 4.4, in connection with the issue of how closely galaxies
trace mass.   

The measured two- through four-point galaxy correlation
functions on scales $\la r_o$ are simply fit by a
scale-invariant clustering hierarchy, a fractal with dimension
$D=3-\gamma =1.23\pm 0.04$. The fractal model does
not fit current ideas on the character of the mass 
distribution, but these ideas are debatable enough
to leave open the possibility that the small-scale clustering
hierarchy is of more than historical interest.  

Synthetic maps constructed by the clustering hierarchy
prescription look reasonably like the Lick galaxy map, provided
the hierarchy is cut off at $hr\sim 20$~Mpc 
(Soniera \&\ Peebles 1978). 
If instead the hierarchy extends to much larger scales, so 
$\xi (r)\propto r^{-\gamma}$ to $\xi\ll 1$, the large scale of 
statistically related fluctuations has a numerically small effect
on the correlation functions but a distinct effect on the 
visual impressions of the synthetic map, making it look
``blotchy.'' Our first check that this is remedied by truncating
the hierarchy was to cut a
``blotchy'' map into strips $\sim 20$~Mpc wide, randomly
reassemble, and repeat in the orthogonal direction. To match 
the evidence for a rather sharp break in the angular two-point
function, we had to arrange the truncation so the spatial
function, $\xi (r)$, rises above the power law at $r\sim r_o$, and
then breaks sharply downward. 

The clustering hierarchy model produces voids of reasonable size
(Vettolani et al. 1985). I don't know how well it might agree
with the striking tendency to smooth walls around voids (de
Lapparent, Geller \&\ Huchra 1986). Rich clusters of galaxies are
not represented in the hierarchy model; one has to picture them as
places where dynamical relaxation has erased the hierarchy. 
I don't know whether a hierarchical model fixed to make realistic
clusters might agree with the Bahcall-Soneira~(1983) scaling of
the cluster richness with the cluster-cluster clustering length.  

\section{How Well do Galaxies Trace Mass?}

The complicated history of ideas on the relation between the
distributions of galaxies and mass has given us a rich suite of
observations and theory that can be used to argue for or against 
the proposition that galaxies are useful mass tracers. 
We do know that, if conventional gravity physics is a good
approximation on the scale of galaxies, most of the mass of a
spiral or large elliptical galaxy is in a dark halo outside
most of the starlight. On this scale starlight certainly is not a
good tracer of mass. But the issue is
whether there are length scales on which galaxies are good
tracers in the sense that the galaxy 
$N$-point correlation functions, or the galaxy-mass cross
correlation functions, are close approximations to the 
mass autocorrelation functions. I begin with a review of results
from the 1970s that, in my reading of the evidence, argue for 
close relations between galaxies and mass. 

\subsection{Suggestions from the Statistics}

I was --- and am --- taken with the following three aspects of
the galaxy and mass $N$-point correlation functions. 

First, the galaxy two-point function, $\xi (r)$, is large and
positive at $r\la r_o$, and quite close to zero but likely
negative at larger separations. The negative part is thought 
to be a result of initial conditions; $\xi (r)$ need not be
negative at any separation $r$. To see this, recall that in one
version of the  Neyman-Scott (1952) prescription all galaxies
are in clusters, and the clusters are placed as a stationary
homogeneous Poisson process. In this construction, the probability
a galaxy is 
found in the volume element $dV$ at distance $r$ from a galaxy is
the sum of $ndV$, the contribution of all the other clusters,
because their positions are statistically independent, and the 
contribution from the cluster to which the galaxy belongs. That
is, $\xi (r)=0$ at $r$ greater than the largest cluster diameter,
and $\xi (r)> 0$ at smaller $r$, from the chance of encountering
a galaxy belonging to the same cluster. It is equally easy to
construct locally compensated density fluctuations: make an
initially smooth galaxy distribution clustered by drawing the
galaxies in regions of space of size $\sim r_o$ into tighter
clumps. This local rearrangement makes $\xi (r)$ positive at
small $r$, and $\xi (r)$ negative, representing anticorrelation,  
at $r\sim r_o$, in such a way that the integral  
\beq
J(r) = \int _0^r\xi (r) d^3r
\eeq
rapidly converges to zero at $r\ga r_o$. 

This rapid convergence is not observed. One might have expected
to have seen it if the galaxy clustering were a result of local
rearrangement, as by explosions (Peebles 1974b). Our folk theorem 
was that, if the strongly nonlinear small-scale clustering grew by
gravity out of a primeval power spectrum that is close to flat,
the mass autocorrelation function would
be positive or zero everywhere. This is because gravitational
flow has shear but no divergence in linear perturbation theory.
In the 1970s this seemed to be a pretty good reason to suspect
that gravity rather than explosions produced the clustering of
the galaxies. Now other evidence for the gravitational
instability picture includes large-scale flows (Courteau, Strauss
\&\ Willick 2000) and the CBR anisotropy (Hu  et al. 2000). 

The galaxies would be expected to have followed the growth of
clustering of the mass if galaxies had been around long enough to
be drawn into clumps with the mass (Peebles 1986; Tegmark \&\
Peebles 1998).  

The second aspect is a numerical coincidence between the
small-scale galaxy correlation 
function and the dark mass halos typical of large
galaxies. The mean number density of galaxies at distance $r\ll
r_o$ from a galaxy (in the sense of the coadded density of
neighbors averaged across a fair sample of galaxies) is 
$n _r=n\xi (r)=n(r_o/r)^\gamma$, where $n$ is the cosmic mean
number density. The  
mean mass density at distance $r\la 100$~kpc from a 
large spiral galaxy is reasonably well approximated as
$\rho _r\sim v_c{}^2/(4\pi G r^2)$, where the circular velocity,
$v_c$, varies only slowly with radius. The cosmic mean mass
density, $\rho$, is represented as
$\Omega H_o^2=8\pi G\rho /3$, where $\Omega $ is the density
parameter (in matter capable of clustering). The ratio is
\beq
{\rho _r\over\rho} \sim {2\over 3\Omega}
\left( v_c\over H_or\right) ^2.
\label{eq:halo}
\eeq
Most of the cosmic luminosity density comes from galaxies like
the Milky Way, where $v_c\sim 200$ km~s$^{-1}$. With this number, 
and at radius $hr=10$~kpc, where the dark halo is becoming dominant,
the ratios of local to cosmic number and mass densities are
\beq
{n_r\over n}\sim 6\times 10^4, \qquad 
{\rho _r\over\rho }\sim {3\times 10^4\over\Omega}.
\label{eq:number}
\eeq
These numbers agree to a factor of two (for the formerly
popular value, $\Omega = 1$, or the current favorite, 
$\Omega\sim 0.25$). 

The hierarchical model (\S 3.2) applied to the mass distribution,
with the fractal dimension of the galaxy distribution, has to
fail on small scales, because the massive halos of galaxies are
not fractal. If the small-scale part of the mass autocorrelation
function, $\xi _{\rho\rho}(r)$, is dominated by massive halos
with the density run in equation~(\ref{eq:halo}), it gives 
\beq
\xi _{\rho\rho}(r) = 
{\pi ^3\over 2}\left(\rho _r\over\rho\right) ^2n_gr^3 
\sim 2\times 10^3,
\eeq
at $hr=10$~kpc, galaxy number density 
$n_g\sim 0.01h^3$~Mpc$^{-3}$, and $\Omega =0.25$. This is a
factor of 30 below $\xi (r)$. As for clusters, one might 
think of massive halos as regions where the hierarchy  
is erased, maybe leaving a signature in the similarities of the
galaxy-galaxy correlation function and the galaxy-mass cross
correlation function in equations~(\ref{eq:halo})
and~(\ref{eq:number}). 

The third aspect is the remarkable regularity of the low
order galaxy spatial correlation functions. The two-point
function, $\xi (r)$, is a good approximation to a power law over
three 
orders of magnitude in the separation, $r$, and five orders of
magnitude in the value of $\xi$ (eq.~[\ref{eq:power_law}]). The 
measured values of the three-point function span some seven
orders of magnitude (eq.~[\ref{eq:zrange}]). The simple model in
equation~(\ref{eq:zetamodel}) fits, to the accuracy of the
measurements, apart from the shape-dependence at the large-scale 
end.  

Are these regularities physically significant? As remarked, I
used to think the gravitational growth of clustering could  
produce mass correlation functions with the same regularities. If
so, it would be good evidence galaxies trace mass. This now 
seems questionable (Ma \&\ Fry 2000a). The issues are reviewed 
in \S 4, but a general point is worth noting. 

If structure on the scale of galaxies and larger grew by gravity, 
and the galaxy correlation functions were not good approximations
to the mass functions, the situation would be curious. The mass
functions, that matter for the operation of gravity, would
not have the  striking regularities of the galaxy functions. The  
mass functions could have regularities of their own, that are 
related to the galaxy functions in some subtly elegant way. Or  
maybe the regularities of the galaxy functions are only
accidents. But both seem contrived. On the face of it, 
the reasonable conclusion would be that the galaxy and 
mass functions are closely related. There are other
considerations, of course. 

\subsection{The Biased Galaxy Formation Picture} 

In this picture for structure formation galaxies are more strongly
clustered than mass. This naturally follows from the adiabatic
cold dark matter (CDM) model for galaxy formation, as will be
discussed, and it offers an elegant way to reconcile the
small relative peculiar velocities of galaxies outside the rich
clusters with the mass density of the Einstein-de~Sitter
model.\footnote{It will be recalled that the Einstein-de~Sitter
cosmological model has negligible curvature of space sections at
fixed world 
time, and a negligible cosmological constant $\Lambda$. The ratio
of the cosmic mean mass density to the Einstein-de~Sitter value
is $\Omega$} The cosmic mean mass density is now thought to be
well below the Einstein-de~Sitter value, but the history of ideas
is worth remembering. I begin with the issue of the mass density.  

Early redshift surveys revealed that the only galaxies with
negative redshifts (corrected for the rotation of the Milky Way)
are members of the Local Group or the Virgo Cluster (Humason,
Mayall \&\ Sandage 1956). If this were representative it would
mean galaxy peculiar velocities\footnote{The peculiar
velocity is measured relative to the homogeneous expansion of the
Hubble flow.} outside rich clusters are no more than a few
hundred km~s$^{-1}$, well below velocities within clusters.
Consistent with this, a statistical analysis of redshifts in the
{\it Reference Catalogue of Bright Galaxies} (de Vaucouleurs \&\
de Vaucouleurs 1964) indicated that the small-scale
one-dimensional rms relative velocity dispersion is only 
about 200 km~s$^{-1}$ (Geller \&\ Peebles 1973). 

Fall (1975) introduced an important application of the relation
between the rms mass peculiar velocity and the gravitational
potential energy measured by the mass autocorrelation function
(Irvine 1961; further discussed by Layzer 1963). Fall
naturally used the galaxy correlation function. He 
found that if $\Omega =1$, as in the Einstein de~Sitter model,
the rms peculiar velocity would have to have to exceed 
1000 km~s$^{-1}$, well above what was suggested by the
observations. 

In the mid 1970s Zwicky's (1933) missing mass problem was under
discussion, albeit muted, and people generally considered the
value of the cosmological density parameter, $\Omega$, a number to
be measured rather than predicted. An example of the former is
the remark in Geller \&\ Peebles (1973), that the relative
velocity dispersion is larger than would be expected from the
mass in the luminous parts of the galaxies, an indication of
missing mass. An example of the latter is Fall's (1975)
conclusion that $\Omega$ likely is well below
unity. He expressed no regret; that was the straightforward
reading of the evidence (Gott et al. 1974). 
I regretted Fall's result, because the scale-invariant
Einstein-de~Sitter model seemed the best way to accommodate the
gravitational growth of a near scale-invariant nonlinear galaxy
clustering hierarchy (Peebles 1974a; Davis, Groth \&\ Peebles
1977). 

It is easier to measure relative velocities of galaxies than 
the peculiar velocities that enter Irvine's (1961) relation. The
rms relative velocities are related to $\Omega$ through the mass
two- and three-point correlation functions (Peebles 1976, which
generalizes Geller \&\ Peebles 1973). A first application gave 
$\Omega = 0.4\pm 0.1$, close enough to unity to keep alive my
hopes for the Einstein-de~Sitter model (Peebles 1979). But the 
result from the application to the Center for Astrophysics (CfA)  
redshift sample (Huchra et al. 1983) was quite inconsistent
with the Einstein-de~Sitter model, if galaxies trace mass 
(Davis \&\ Peebles 1983). 

We now know the peculiar velocity of the Local Group is large,
$\sim 600$ km~s$^{-1}$, but that is thought to be a result of the
large coherence length of the mass fluctuations, not large
$\Omega$. Also, the CfA relative velocity
dispersion is biased low, because rich clusters of galaxies are
under-represented (Marzke et al. 1995). But the mass within the
rich clusters certainly is well below the Einstein-de~Sitter
value; it is the low relative velocity dispersion outside
clusters that would be so difficult to reconcile with the
Einstein-de~Sitter mass, if it were concentrated with the field
galaxies, as was pretty clearly evident in the early 1980s. 

In the early 1980s many had accepted the inflation picture of the
very early universe as a compelling argument for the
Einstein-de~Sitter model. That led to an intense interest in the
nature and amount of the missing mass, under its new name, dark
matter. Inflation  and dark matter were the leading talks at the
1982 Texas Symposium on Relativistic Astrophysics (Guth 1984;
Pagels 1984). A crisis was the evidence that the mass density is
less than the critical Einstein-de~Sitter value.  

A resolution, biased galaxy formation, grew out of an
excellent remark by Kaiser (1984). The two-point correlation
function of positions of the rich clusters of galaxies is much
larger than the galaxy correlation function 
(Peebles \&\ Hauser 1974; Bahcall \&\ Soneira 1983).
Kaiser (1984) showed that if clusters were the rare peaks in the
mass distribution from the gravitational growth of
clustering out of Gaussian fluctuations with a broad
coherence length, clusters would be more strongly clustered 
than mass. Kaiser (1986) and Bardeen (1986) extended the 
thought: if most galaxies formed at less rare density peaks, they
would be less clustered than the clusters, but more strongly
clustered than the mass, as wanted.  

These considerations led Davis et al. (1985) to explore biased
galaxy formation in numerical simulations of the CDM
model. Giant galaxies would 
form preferentially in high density regions, in tight
concentrations. The assembly of less massive galaxies would tend
to be completed later, in less dense regions, in looser
concentrations. Voids between the concentrations of ordinary
$L\sim L_\ast$ galaxies\footnote{$L_\ast$ is the luminosity at the 
knee of the luminosity function, and the characteristic
luminosity of the galaxies that produce most of the starlight}
would contain most of the mass. The void mass would be seeded for
galaxy formation, but seeds for $L\sim L_\ast$ galaxies in voids
would tend to germinate late, under conditions that could be
unfavorable for the development of observable galaxies. 

Biased galaxy formation agrees with the observation that the most
massive galaxies, such as the cD galaxies that prefer to be 
in clusters, are more strongly clustered than $L\sim L_\ast$
galaxies (Hamilton 1988; Valotto \&\ Lambas 1997), though the
effect is small (Szapudi et al. 2000a). Biasing is
not so clearly consistent with the similar distributions of
bright and faint galaxies outside clusters, an early example of
which is the strikingly similar maps of bright and faint 
galaxies in the CfA sample (Davis et al. 1982, Figs.~2a and~2d).
This phenomenon led me to learn to like a low density universe 
(Peebles 1984, 1986). Other early reasons to consider a low
density spatially flat universe are discussed by Turner, Steigman
\&\ Krauss (1984), Vittorio \&\ Silk (1985), and Efstathiou,
Sutherland \&\ Maddox (1990).  

Recent advances in applications of the cosmological tests
indicate the density parameter in matter capable of clustering is
(Hu et al. 2000; Bahcall et al. 2000) 
\beq
\Omega = 0.25\pm 0.1. 
\label{eq:Omega}
\eeq
This would mean there is no need to sequester mass in the voids:
galaxy relative peculiar velocities outside clusters, on scales 
\beq
100\hbox{ kpc}\la hr\la 10\hbox{ Mpc},
\label{eq:xx}
\eeq
are consistent with the assumption that galaxies trace mass.

Though biased galaxy formation is no longer thought to be needed,
numerical simulations of the CDM model still predict the presence
of low mass halos between the concentrations of dark matter halos
that would seem to be suitable homes for $L\sim L_\ast$ galaxies. 
This does not violate equation~(\ref{eq:xx}), because little mass
is involved, but it is quite at odds with the observations. A
vivid illustration comes from the extensions of the CfA redshift
survey (de Lapparent, Geller \&\ Huchra 1986; Thorstensen et 
al. 1995), that add many low luminosity galaxies to the low
redshift part of the first CfA survey. These faint galaxies more
sharply define the voids, rather than spilling into them in the
way suggested by CDM simulations. This applies to a broad
range of objects, including dwarf and irregular galaxies,
star-forming galaxies, low surface brightness galaxies, and high
surface density gas clouds. The observational literature is
reviewed in Peebles (2001a). I consider it a serious crisis for
the low density CDM model. 

\subsection{Correlation Dynamics}

Since the use of correlation functions to describe the galaxy 
distribution was inspired in part by the example of nonideal
gases, it was natural to consider the application of a dynamical
method from the analysis of nonideal gases, the BBGKY hierarchy
of relations among correlation functions in position and
momentum. This was first done by Saslaw (1972), Fall \&\ 
Saslaw (1976), Fall \&\ Severne (1976) and Davis \&\ Peebles 
(1977). Perhaps the key lesson from the last reference is that
the problem is exceedingly difficult with available methods. 

The sensible response is to develop the perturbation theory of
the growth of initially small departures from homogeneity. Fry
(1984) led this approach. Recent applications 
(Frieman \&\ Gazta\~naga 1999; Feldman et al. 2001) show that if 
(1) the galaxy correlation functions are good approximations to the
mass functions on the scale of weakly nonlinear departures from
homogeneity, and (2) the clustering grew out of Gaussian initial
fluctuations, then perturbation theory is in remarkably good
agreement with the galaxy three-point functions derived from
optical and infrared catalogs. The first assumption agrees with
other evidence for a low density universe in which galaxies trace
mass on scales $r\ga r_o$ (eq.~[\ref{eq:Omega}]). The
situation on smaller scales is discussed next.  

\subsection{The Halo Clustering Model}

The Neyman-Scott prescription (\S 3.2) places all
galaxies in clusters. The probability a galaxy is placed in the
volume element $d^3r$ at position $\vec r$ relative to the
cluster center is 
\beq
dP \propto \rho (r)\, d^3r.
\label{eq:NS}
\eeq
The constant of proportionality may be a random number,
independently assigned to each cluster. If the cluster centers
are uniformly  distributed one can use the free function, the
cluster number density run $\rho (r)$, to fit the small-scale galaxy
correlation function, $\xi (r)$. One can anticorrelate the
cluster positions to fit the weak anticorrelation of galaxies at
large separations. Variants of the Neyman-Scott prescription
appear in McClelland \&\ Silk (1978), Scargle (1981), Scherrer
\&\ Bertschinger (1991), and in the halo model 
(Sheth \&\ Jain 1997; Ma \&\ Fry 2000a,
2000b; Seljak 2000; Peacock \&\ Smith 2000). The halo model is
suggested by numerical simulations of the CDM model for structure 
formation. The ideas in earlier references are at most only
loosely related; the prescription is broadly appealing. 

I used to think that adjusting the prescription to
fit the higher order galaxy correlation functions is problematic.
If the cluster number density run were taken 
to be a power law, $\rho (r)\propto r^{-\epsilon}$, to fit the
small-scale galaxy two-point function, the predicted three- and
four-point functions would be quite unacceptable 
(Peebles \&\ Groth 1975; McClelland \&\ Silk 1978; LSS \S 61). 
Ma \&\ Fry (2000a) present a numerical demonstration. With two
free functions, $\rho (r)$ and the frequency distribution of
cluster richness as a function of cluster mass, one can 
fit the galaxy two-point function and the scaling of the
three-point function $\zeta$ with triangle size at fixed triangle
shape. That would leave the problem of fitting the variation
of $\zeta$ with triangle shape (eq.~[\ref{eq:zetamodel}]), and of
fitting the galaxy four-point function. But the halo cluster
model takes this approach.

Generalizations of the Neyman-Scott prescription to fit the higher
order correlation functions were considered by Neyman, Scott \&\
Shane (1956) --- who had in mind the evidence for hierarchical
clustering --- and McClelland \&\ Silk (1978). The present
state of ideas, in the context of the halo model, is reviewed 
in Ma \&\ Fry (2000b) and Scoccimarro et al. (2001). To my
mind the main point of principle emerging from these 
considerations is that the CDM model seems to predict that 
at $r\la r_o$ the
mass three-point autocorrelation function is quite
different from the galaxy three-point function: galaxies
are not good tracers of nonlinear mass fluctuations.

\subsection{So How Well Do Galaxies Trace Mass?}

The theoretical situation seems clear: the adiabatic CDM model
predicts significant differences between the small-scale
distributions of galaxies and mass, in two aspects. First, biased
galaxy formation is a natural consequence of the model, in
particular the prediction that objects with relatively low mass
dark halos spread into the voids defined by the halos of 
$L\sim L_\ast$ galaxies. Second, the CDM halo model, with a
suitable prescription for the assignment of numbers of galaxies
to halos, predicts a fit to the galaxy three-point function at
$r<r_o$ that is quite different from the mass three-point
function. 

I distrust the first point because it doesn't agree with the
observations: all known galaxy types avoid the voids. I 
dislike the second point because it reminds me of epicycles; I
regret the loss of simplicity of a scale-invariant fractal. 

The evidence that we live at a special time in the evolution of
the universe, at the transition to $\Lambda$-dominated expansion,
is a cautionary demonstration that Nature does not always choose 
the apparently simplest way. Are the simple regularities of the
galaxy correlation functions at $r\la r_o$ physically
significant, or only a sequence of accidents? The adiabatic CDM
model led us to this conundrum; should we trust the model?  

The dramatic success of the low density cosmologically flat CDM 
model in correlating the measurements of the CBR anisotropy with 
astrophysically reasonable parameters (Hu et al. 2000) argues
this is a good approximation to how structure started forming, at
redshift $z\sim 1000$, on the length scales probed by the CBR
measurements. The CDM model also 
successfully coordinates observations of cosmic structure on
the smaller scales of superclusters down to groups of 
galaxies (Bahcall et al. 2000). But Sellwood \&\ Kosowsky (2000)
list deep challenges to the CDM model on the scale of galaxies.
I would add the issue of the epoch of galaxy formation 
(Peebles 2001b), and, on intermediate length scales, the challenge
of the void phenomenon.   

In short, I can see good arguments for and against the CDM
prediction that the galaxy correlation functions are not useful
approximations to the mass functions on scales $\la r_o$.

\section{Other Histories}

The $N$-point correlation functions and related statistics that
are the subject of this review certainly are not always
the most useful. Nearest neighbor statistics are a
better probe for voids. The topology of large-scale structure
(Gott et al. 1989) reveals effects not seen in the correlation
functions. And one should not underestimate the importance of
visual comparisons of data and synthetic maps 
(Scott, Shane \&\ Swanson 1954; Kiang 1967; Soneira 
\&\ Peebles 1978). It would be fascinating to see a
map at the Lick depth based on the CDM halo model.

I have touched on some aspects of the history of ideas of how 
the galaxies formed, and the establishment of the gravitational
instability picture in the form of the low density CDM model.
Many other ideas for structure formation were considered in the
last two decades; they employ excellent physics that could
reappear in theories of the astrophysics. The universe is 
complicated; maybe structure formation is too. Explosions caused
by superconducting cosmic strings (Ostriker, Thompson \&\ Witten
1986) could make admirable voids, for example, provided they were
subdominant enough not to be detectable on the scale of the
present weakly nonlinear clustering.  

\acknowledgments
I am grateful to Marc Davis, Jim Fry, Ed Groth, Chung-Pei Ma,
Rom\'an Scoccimarro and Uros Seljak for helpful discussions.  
This work was supported in part by the USA National Science
Foundation.

\end{document}